\documentclass[iop]{emulateapj}
\usepackage{graphicx}
\usepackage{color}
\usepackage[dvipsnames]{xcolor}
\usepackage{longtable}
\usepackage{mdwlist}
\usepackage{color}
\usepackage{enumitem}
\usepackage{lipsum}
\slugcomment{version \today: fm}
\shorttitle{The dragon's lair}
\shortauthors{F. Massaro et al.}

\begin{document}
\title{Dragon's Lair: \\ on the large-scale environment of BL Lac objects}
\author{F. Massaro\altaffilmark{1,2,3,4}, A. Capetti\altaffilmark{2}, A. Paggi\altaffilmark{1,2,3}, R. D. Baldi\altaffilmark{1,5}, \\ A. Tramacere\altaffilmark{6}, I. Pillitteri\altaffilmark{7} \& R. Campana\altaffilmark{8}}
\altaffiltext{1}{Dipartimento di Fisica, Universit\`a degli Studi di Torino, via Pietro Giuria 1, I-10125 Torino, Italy.}
\altaffiltext{2}{INAF-Osservatorio Astrofisico di Torino, via Osservatorio 20, 10025 Pino Torinese, Italy.}
\altaffiltext{3}{Istituto Nazionale di Fisica Nucleare, Sezione di Torino, I- 10125 Torino, Italy.}
\altaffiltext{4}{Consorzio Interuniversitario per la Fisica Spaziale, via Pietro Giuria 1, I-10125 Torino, Italy.}
\altaffiltext{5}{Department of Physics and Astronomy, University of Southampton, Highfield, SO17 1BJ, UK.}
\altaffiltext{6}{University of Geneva, Chemin d'Ecogia 16, Versoix, CH-1290, Switzerland.}
\altaffiltext{7}{INAF-Osservatorio Astronomico di Palermo G.S. Vaiana, Piazza del Parlamento 1, 90134, Italy.}
\altaffiltext{8}{INAF/OAS, via Piero Gobetti 101, I-40129, Bologna, Italy.}

\begin{abstract} 
The most elusive and extreme sub-class of active galactic nuclei (AGNs), known as BL Lac objects, shows features that can only be explained as the result of relativistic effects occurring in jets pointing at a small angle with respect to the line of sight. A long standing issue is the identification of the BL Lac parent population, having jets oriented at larger angles. According to the ``unification scenario'' of AGNs, radio galaxies with low luminosity and edge-darkened radio morphology are the most promising candidates to be the parent population of BL Lacs. Here we compare the large-scale environment, an orientation independent property, of well-defined samples of BL Lacs with samples of radio-galaxies all lying in the local Universe. Our study reveals that BL Lacs and radio galaxies live in significantly different environments, challenging predictions of the unification scenario. We propose a solution to this problem proving that large-scale environments of BL Lacs is statistically consistent with that of compact radio-sources, known as FR\,0s, sharing similar properties. This implies that highly relativistic jets are ubiquitous and are the natural outcome of the accretion of gas into the deep gravitational potential well produced by supermassive black holes.
\end{abstract}

\section{Introduction}
\label{sec:intro} 
Since the early 70's, extended radio galaxies were divided in two main types based on their radio morphology, distinguishing between edge-darkened (FR\,I type) and edge-brightened (FR\,II type) sources \citep{fanaroff74}. For decades this dichotomy, was linked to their radio power and their large-scale environments having FR\,Is that generally inhabit galaxy-rich environments, being members of groups or galaxy clusters, while FR\,IIs live more isolated \citep[][see e.g.,]{zirbel97}. Radio galaxies were also classified on the basis of their optical spectra \citep{hine79}, distinguishing between high and low excitation radio galaxies (HERGs and LERGs, respectively). While LERGs can show both FR\,I or FR\,II radio morphology \citep[see e.g.,][]{laing94} HERGs appear to be, almost exclusively, FR\,IIs \citep[see also][]{heckman14}.

Furthermore, it is becoming clear that the majority of low redshift radio galaxies are compact sources. These, known as FR\,0s, have with typical sizes $\lesssim 10 kpc$ and are characterized by a LERG spectrum \citep{baldi15} and as recently showed tend to live in poorer environments with respect to extended radio sources \citep{capetti20}.

On the other hand, BL Lac objects (hereinafter BZBs) are now recognized as the most extreme class of AGNs. Emitting from radio to TeV energies, they constitute the largest population of gamma-ray sources \citep{abdollahi20} and show several peculiar observational properties including: flat radio spectra \citep{healey07,massaro13}, apparent superluminal motions \citep{lister13}, extreme variability up to TeV energies \citep{aharonian07}, high radio-to-optical polarization \citep{pavlidou14}, peculiar mid-infrared colors \citep{massaro11} and featureless optical spectra, with only weak emission/absorption features \citep{stickel91}. At the Pittsburgh Conference in 1978 Blandford and Rees proposed to interpret all these features as non-thermal emission arising from particles flowing in a relativistic jet observed at a small angle with respect to the line of sight. 

According to the ``unification scenario'' of radio-loud AGNs, at zeroth order, all jetted AGNs are intrinsically the same but they appear diverse due to different orientations with respect to the line of sight \citep{urry95}. This idea immediately prompted the quest for the identification of misaligned BZBs. Among radio-loud AGNs, radio galaxies, mainly those belonging to the FR\,I radio class \citep{fanaroff74}, having low luminosity and an edge-darkened radio morphology, were naturally identified as the BZB parent population, since they also produce relativistic jets, extending up to hundreds of kiloparsec scales and lack broad emission lines.

There is a vast literature of tests on the validity of this unification scenario, in particular, those performed on the basis of the study of the large-scale environments \citep[see e.g.,][]{villarroel14,zou19}, an orientation independent property of AGNs \citep{antonucci85}. Here we carry out a statistical environmental test of the ``unification scenario'', comparing a selected sample of BZBs with both (i) FR\,Is and (ii) LERGs, aiming at verifying if radio galaxies and BZBs inhabit the same galaxy-rich large-scale environments.  
 
We adopt cgs units for numerical results and we assume a flat cosmology with $H_0=69.6$ km s$^{-1}$ Mpc$^{-1}$, $\Omega_\mathrm{M}=0.286$ and $\Omega_\mathrm{\Lambda}=0.714$ \citep{bennett14}, unless otherwise stated, as  adopted in previous analyses \citep[see][hereinafter M19 and M20, respectively]{massaro19,massaro20}.
\begin{figure*}[!th]
\begin{center}
\includegraphics[height=6.2cm,width=8.4cm,angle=0]{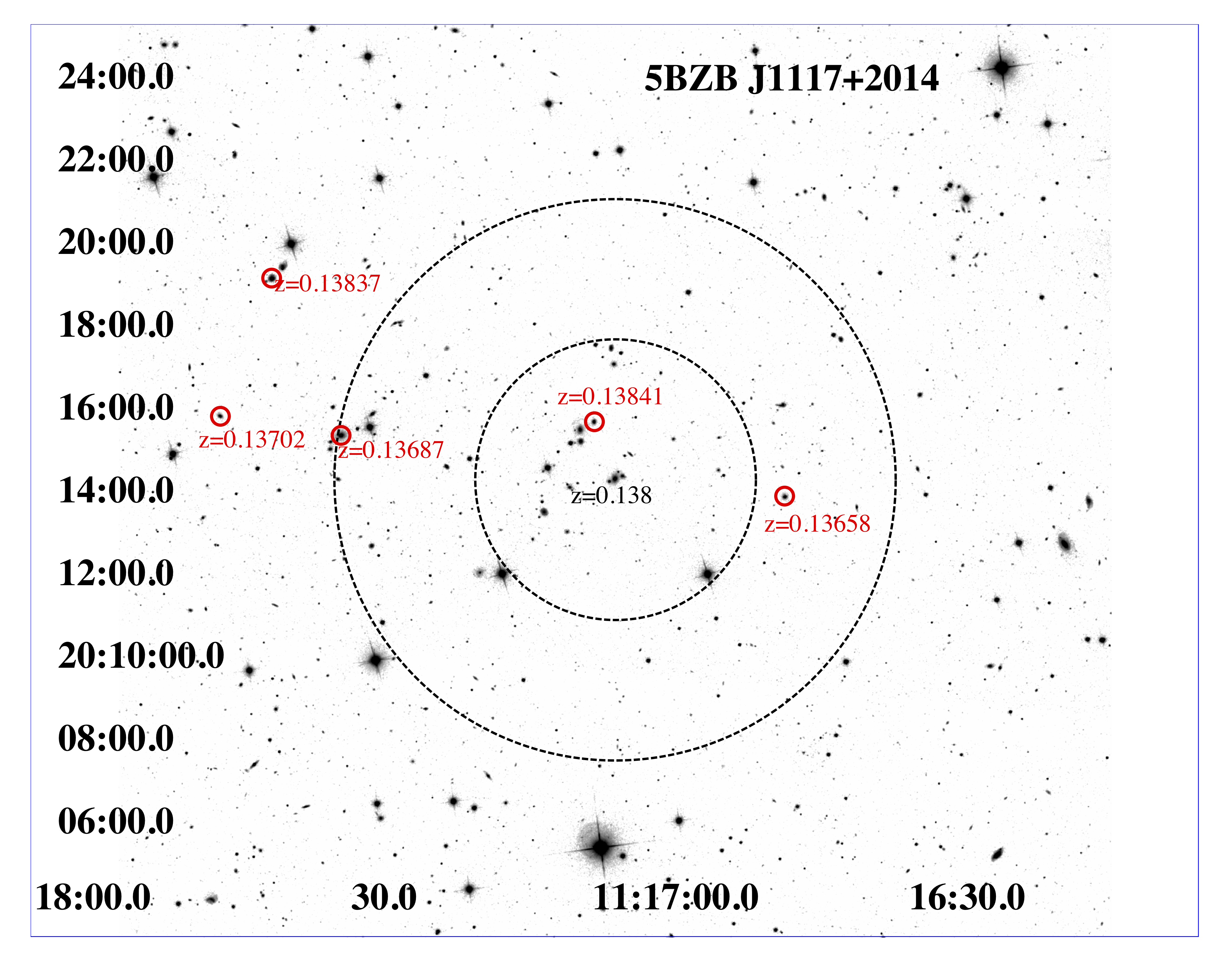}
\includegraphics[height=6.2cm,width=8.4cm,angle=0]{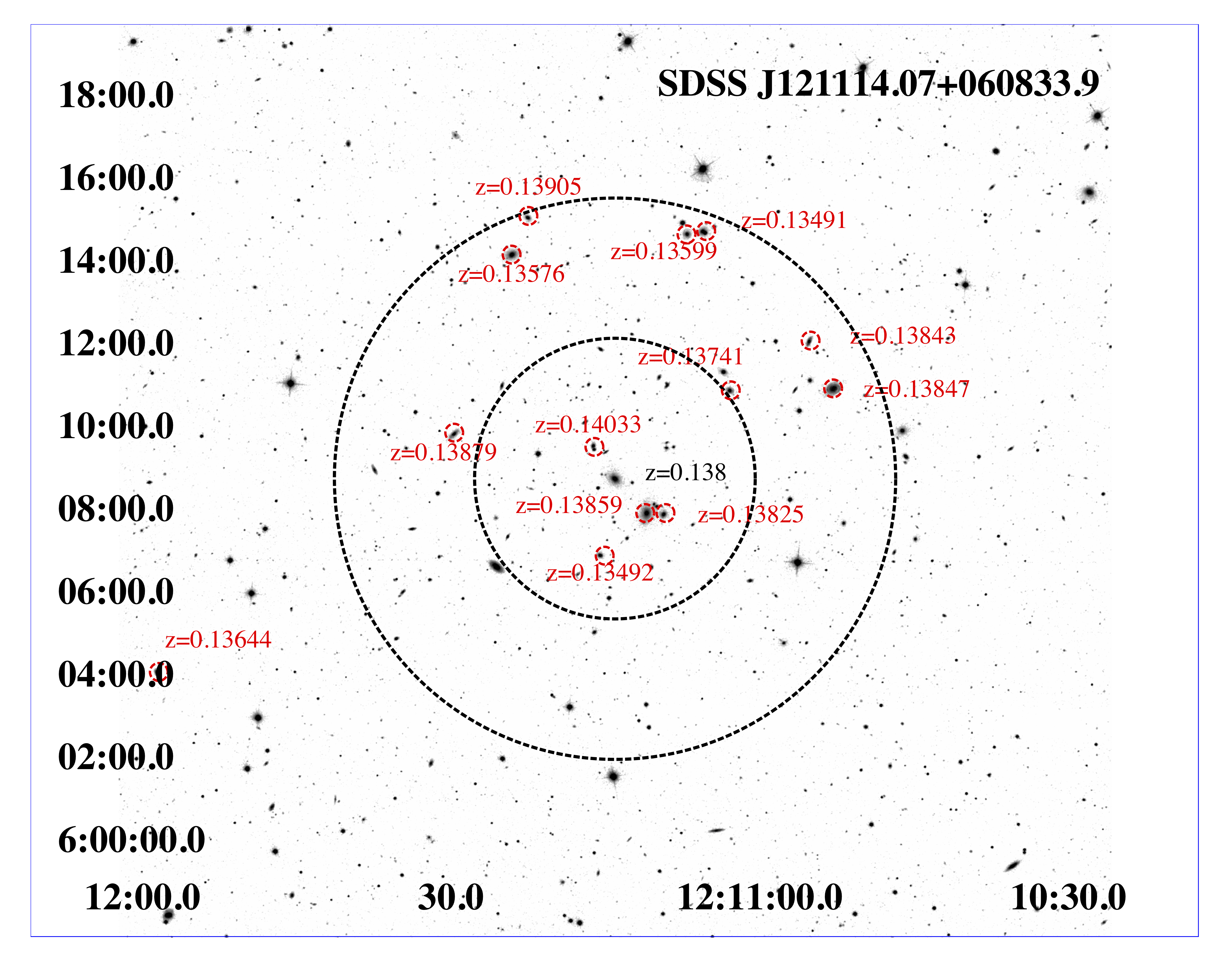}
\end{center}
\caption{{\it Left)} The $R$-band SDSS image of the field surrounding 5BZB J1117+2024 centered on its position. Two lack circles have radius of 500\,kpc and 1\,Mpc, respectively, computed at the central source redshift. All cosmological neighbors are marked with a red circle and have their $z_\mathrm{src}<$ reported. In our sample 5BZB J1117+2024 has the largest number of cosmological neighborhoods within 2\,Mpc. {\it Right)} Same as left panel for the FR\,I SDSS J121114.07+0608339 at the same redshift of 5BZB J1117+2024, both reported close to their positions in black. It is quite evident as SDSS J121114.07+0608339 has a large-scale environment richer of galaxies than 5BZB J1117+2024. In both figures cosmological neighbors are brighter than 17.8 magnitudes in the $R$ band (i.e.. the SDSS threshold to select spectroscopic targets.}
\label{fig:sky}
\end{figure*}

\section{Sample selection}
\label{sec:samples} 
We combined sources listed in the FR\,ICAT with those of the sFRICAT sample for a total of 209 FR\,Is \citep{capetti17a}. The former sample lists FR\,Is at redshifts $z_\mathrm{src}\leq$0.15$\footnote{$z_\mathrm{src}$ indicates the source redshift while $z_\mathrm{cl}$ that of a possible nearby galaxy group/cluster.}$, selected to have a radio structure extending beyond 30 kpc, measured from the location of the host galaxy as seen in the optical band, while the latter includes 14 FR\,Is with radio emission between 10 and 30 kpc and $z_{src}\leq$0.05. Then we also considered 101 FR\,IIs, in the same redshift range and all classified as LERGs, collected out of the FR\,IICAT \citep{capetti17b} to obtain a sample of 310 LERGs.  Both FR\,ICAT and FR\,IICAT are based on data available in the Sloan Digital Sky Survey \citep[see e.g.,][]{ahn12} and the Faint Images of the Radio Sky at Twenty-cm survey \citep[FIRST][]{white97}. 

In our analysis we also performed a comparison with 108 FR\,0 radio galaxies selected in Baldi et al. (2018). These are all radio galaxies with (i) $z_\mathrm{src}<$0.05 being (ii) optically classified as LERG, with (iii) a radio flux density at 1.4 GHz in the FIRST survey above 5 mJy and (iii) lacking extended radio emission beyond a few kpc.

For BZBs we selected only those lying in the same SDSS central footprint and having a firm redshift estimate at $z_\mathrm{src}\leq$0.15, all out of the 5$^{th}$ release of the Roma-BZCAT \citep{massaro15} for a total of 11 sources. Then we also added 3 more BZBs lying at $z_\mathrm{src}<$0.15 that were recently discovered thanks to our optical spectroscopic follow up campaign of low energy counterparts for the unidentified $\gamma$-ray sources \citep{massaro12a,massaro16,demenezes19,pena20}. Thus the final sample of BZBs considered in our analysis lists 14 in the same redshift bin of radio galaxies. 

For a comparison with literature results we also considered BL Lacs-galaxy dominated (hereinafter BZGs) listed in the Roma-BZCAT. Adopting the same criteria as for the BZB selection we extracted 41 BZGs lying between 0.02$<z<$0.15 and with 14 out of 41 being associated with a $\gamma$-ray source \citep{abdollahi20}. BZGs are radio sources  whose multifrequency emission exhibits some properties of BL Lacs but appears dominated by the host galaxy contribution, in particular in the optical-ultraviolet energy range. It is not yet clear if BZGs are all genuine BZBs, in a quiescent state given the high variability that BL Lacs show, or are moderately bright AGNs whose non-thermal emission does not present evidence for relativistic beaming \citep[see also][]{massaro12b}.  
\begin{figure*}[!th]
\begin{center}
\includegraphics[height=6.2cm,width=8.4cm,angle=0]{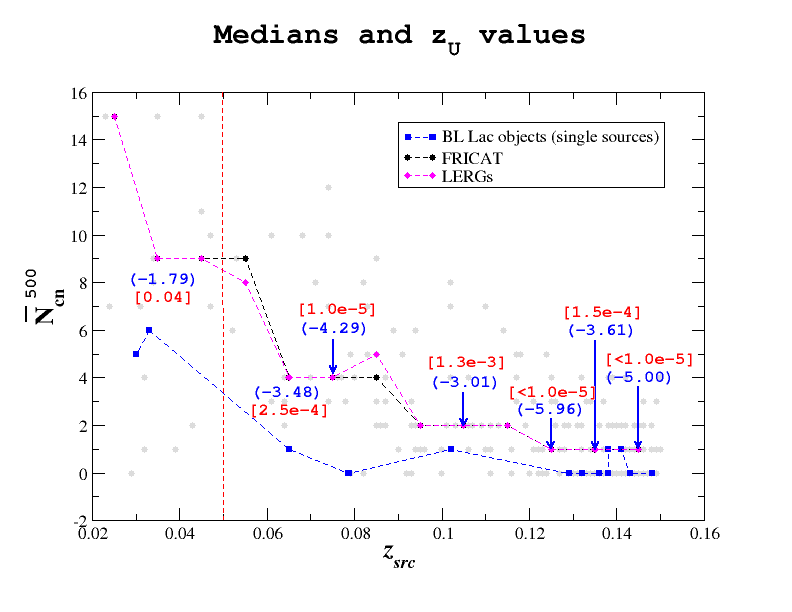}
\includegraphics[height=6.2cm,width=8.4cm,angle=0]{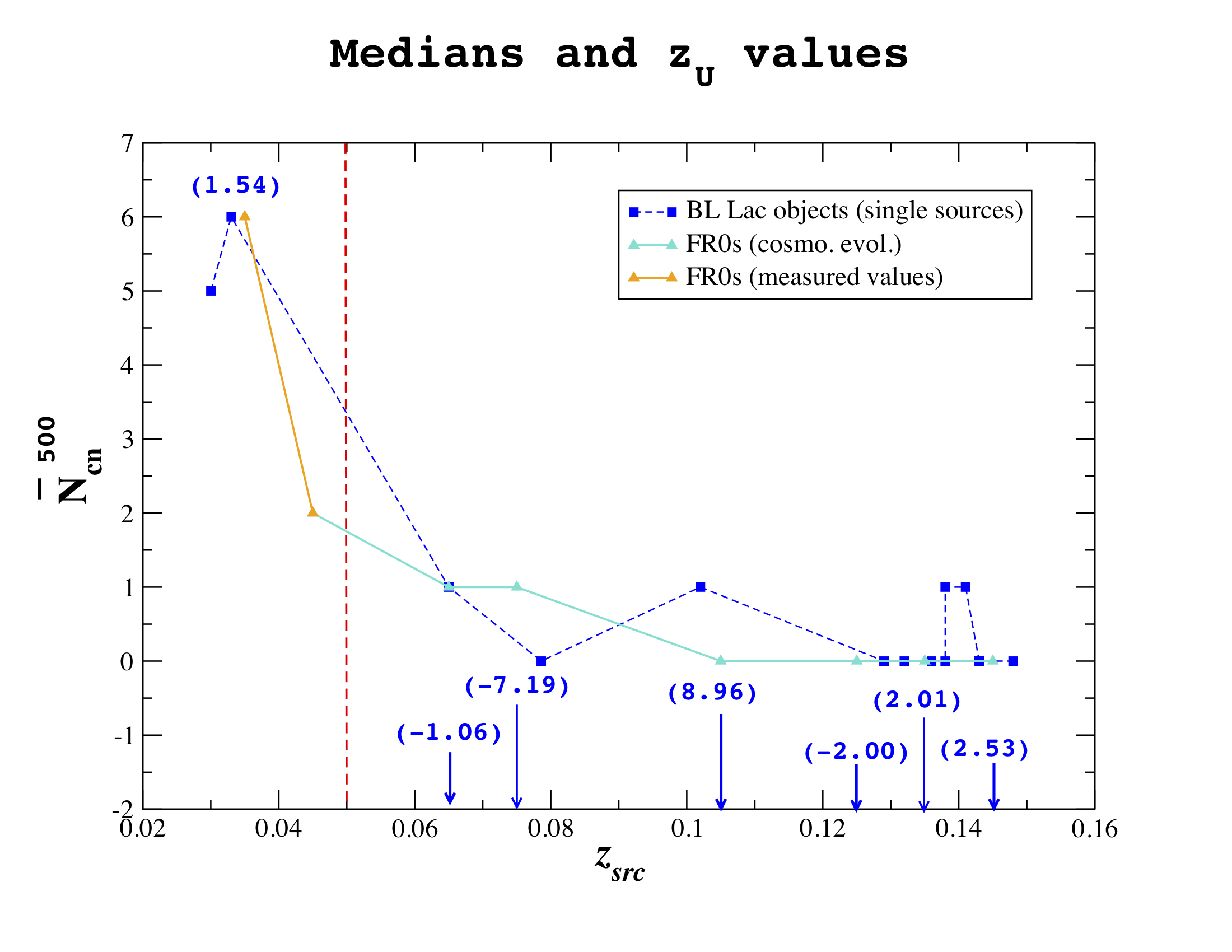}
\end{center}
\caption{{\it Left)} Medians of ${N}^{500}_{cn}$ for FR\,Is (black circles) and LERGs (magenta diamonds) per redshift bins of 0.01 size. Blue squares corresponds to the values for single BZBs at $z_\mathrm{src}<$0.15 Blue numbers reported in parenthesis close to each median of LERGs correspond to values computed for the $z_U$ normalized variable of the Mann-Whitney $U$ rank test performed between BZBs and LERGs (see \S.~\ref{sec:environment} for details) while red numbers correspond to the p-values. As shown BZBs are distributed systematically below all medians of both FR\,Is and LERGs thus indicating that they inhabit less galaxy-rich large-scale environments. All grey circles shown in the background correspond to the single values of ${N}^{500}_{cn}$ for all considered LERGs. {\it Right} Same as left panel but reporting the comparison between BZBs and FR\,0s. Orange triangles are measured medians below $z_\mathrm{src}=$0.05 while cyan triangles refer to simulated medians between 0.05$<z<$0.15 for each bin where there is at least one BZB. It is remarkable the agreement between both measured and estimated/simulated medians of the FR\,0 population and that of BZBs, indicating that the former could be the parent population of BZBs. Vales of the normalized $z_U$ variable are also reported.}
\label{fig:stat}
\end{figure*}

\section{Investigating large-scale environments}
\label{sec:environment} 
The comparison between large-scale environments of BZB, FR\,I and LERG samples, all in the SDSS central footprint, is carried out adopting the same procedure of M19 and M20. We used the number of cosmological neighbors to estimate the environmental richness. These are defined as all optical sources with SDSS magnitude flags indicating a galaxy-type object and having a spectroscopic $z$ with $\Delta\,z=|z_\mathrm{src}-z|\leq$0.005, thus corresponding to the maximum velocity dispersion in groups and clusters of galaxies \citep[see][e.g.,]{berlind06}. We indicate their number as $N_{cn}^{500}$ and $N_{cn}^{2000}$, for those lying within 500\,kpc and 2\,Mpc from the central source, respectively. In Table~\ref{tab:main} all parameters estimated for each sample analyzed here. In Figure~\ref{fig:sky} we show the $R$ band optical image of the field around one BZB and one FR\,I in our samples with all cosmological neighbors highlighted. 

As previously carried out to compare large-scale environments of two different classes we performed the following statistical tests. The first is based on $\bar{N}_{cn}^{500}$, i.e., medians of the $N_{cn}^{500}$ distribution, while the second applying the Mann-Whitney $U$ rank. To avoid cosmological effects both tests performed in each redshift bin of 0.01 size (see M20 for details). 

We also compared the environment of FR\,0s with that of BZBs. However, since the FR\,0 sample is limited to $z=$0.05, we simulate how galaxy over-density surrounding FR\,0s would be detected if they lie at larger redshift adopting the same strategy of Capetti et al. (2020). Thus assuming that large-scale environments of FR\,0s does not evolve in the redshift range between 0.05 and 0.15, we computed the absolute magnitude in the $R$ band of all cosmological neighbors and, maintaining their intrinsic power, we rescaled it at larger distances, i.e. in all redshift bins where there is at least one BZB. We also recomputed accordingly radii of 500\,kpc and 2\,Mpc in each redshift bin. We measured the number of cosmological neighbors with rescaled apparent magnitude $m_r$ brighter than 17.8 corresponding to the SDSS criterion to select spectroscopic targets. These simulations allow us to measure expected medians of cosmological neighbors within 500\,kpc and 2\,Mpc circles for FR\,0s up to $z=$0.15. 

These simulations were tested over the FR\,I samples. Under the same assumptions previously described, we computed median values of all cosmological neighbors surrounding simulated FR\,Is in all redshift bins up to $z=$0.15 where there is at least one BZB and we found a perfect agreement with the observed values being 4 at $z=$0.065, 3 at $z=$0.075 and 1 at $z=$0.105, 0.125 and 0.135, as reported in the following.

\section{Results}
\label{sec:results} 
The median values $\bar{N}_{cn}^{500}$ for both samples of FR\,Is and LERGs are shown in Fig.~\ref{fig:stat}. It is clear that the measured values of $N_{cn}^{500}$ for all 14 BZBs lying at $z_\mathrm{src}<=$0.15 lie all systematically below the $\bar{N}_{cn}^{500}$ of both radio galaxy sample. Then comparing $\bar{N}_{cn}^{2000}$ the situation is in agreement with previous results with only 3 out of 14 BZBs for which $\bar{N}_{cn}^{2000}$ is marginally consistent with that of radio galaxies (i.e., FR\,Is and LERGs). 

In Fig.~\ref{fig:stat} we show, above each $N_{cn}$ median value of the two RG distributions of the normalized $z_U$ variable computed for the Mann-Whitney $U$ test when comparing BZBs with LERGs. This is again systematically negative and not consistent with zero within more than a 3 $\sigma$ level of confidence, with the only exception of a single bin between $z=0.03$ and $z=0.04$. The latter statistical tests were performed in each redshift bin thus grouping BZBs as LERGs. 

Both statistical tests allows us to claim that the hypothesis that large-scale environments of BZBs and that of FR\,Is and/or LERGs are similar can be rejected with high level of confidence, with a chance probability of 10$^{-4}$ for the median test. This implies that FR\,Is or more in general LERGs inhabit richer large-scale environments than BZBs and cannot be their ``parent'' population as predicted by the AGN unification scenario. 

Finally, we compared large-scale environments of BZBs and FR\,0s where measured values of $\bar{N}_{cn}^{500}$ and $\bar{N}_{cn}^{2000}$ of FR\,0s are only available at $z_\mathrm{src}<$0.05. However, adopting simulations previously described (see \S~\ref{sec:environment}) we ``extrapolated'' the behavior of FR\,0 environments up to $z=$0.15. As shown in Fig.~\ref{fig:stat} values of $N_{cn}^{500}$ for BZBs and FR\,0s, both measured (i.e., below $z=$0.05) and extrapolated up to $z=$0.15 with the median test and or using the $z_U$ normalized variable appear indistinguishable.

\section{Comparison with the literature}
\label{sec:compare} 
The largest fraction of all analyses carried out to date on BZB large-scale environments are mainly focused on single sources and small samples \citep[see e.g.,][]{arp70,disney74,craine75,stickel91,torreszafra18,rovero16}, with several being based on photometric companion galaxies in their neighborhoods \citep{falomo90,falomo93a,falomo93b,wurtz93,pesce94,pesce95,falomo95,muriel15}. However results appear to be still controversial and contradictory.

Then in 2016 the comparison between BZBs listed in the Roma-BZCAT and sources belonging to the catalog of galaxy clusters and groups of Merch\`an \& Zandivarez (2005) was presented \citep{muriel16}. This is the first statistical analysis over a large sample of BL Lacs. Muriel (2016) found that 121 blazars appear to be associated with sources listed in the cluster catalog, these are classified as: 24 BZBs, 96 BZGs and 1 BZUs according to the Roma-BZCAT. Restricting the analysis to redshifts below $\sim$0.2, where the cluster catalog is less incomplete the number of spatial coincidences decreases to 78. Taking into account the contamination by spurious groups/clusters of galaxies only 43$\pm$5\% of all BL Lac objects, including BZGs, lie in groups of three or more members, where the expected fraction computed for a random sample of galaxies, having the same redshift distribution, is 19.3$\pm$0.1\%. Muriel et al. (2016) also applied a correction factor due to the redshift incompleteness of the algorithm used to create the galaxy cluster/group catalog then claiming a BZB fraction in groups of $\sim$67$\pm$8\% for all 78 sources.  

According to the Roma-BZCAT, BZGs are not ``genuine'' blazars, but could be moderately bright AGNs whose non-thermal emission does not present evidences for relativistic beaming and/or are misclassified sources. Thus we analyzed BZGs  analyzed separately from BZBs.
\begin{figure*}[!th]
\begin{center}
\includegraphics[height=6.2cm,width=8.4cm,angle=0]{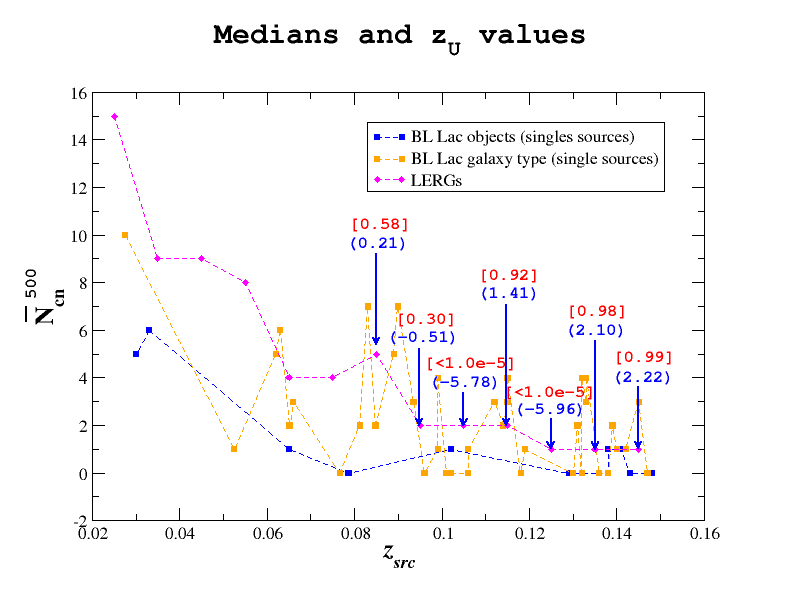}
\includegraphics[height=6.4cm,width=9.4cm,angle=0]{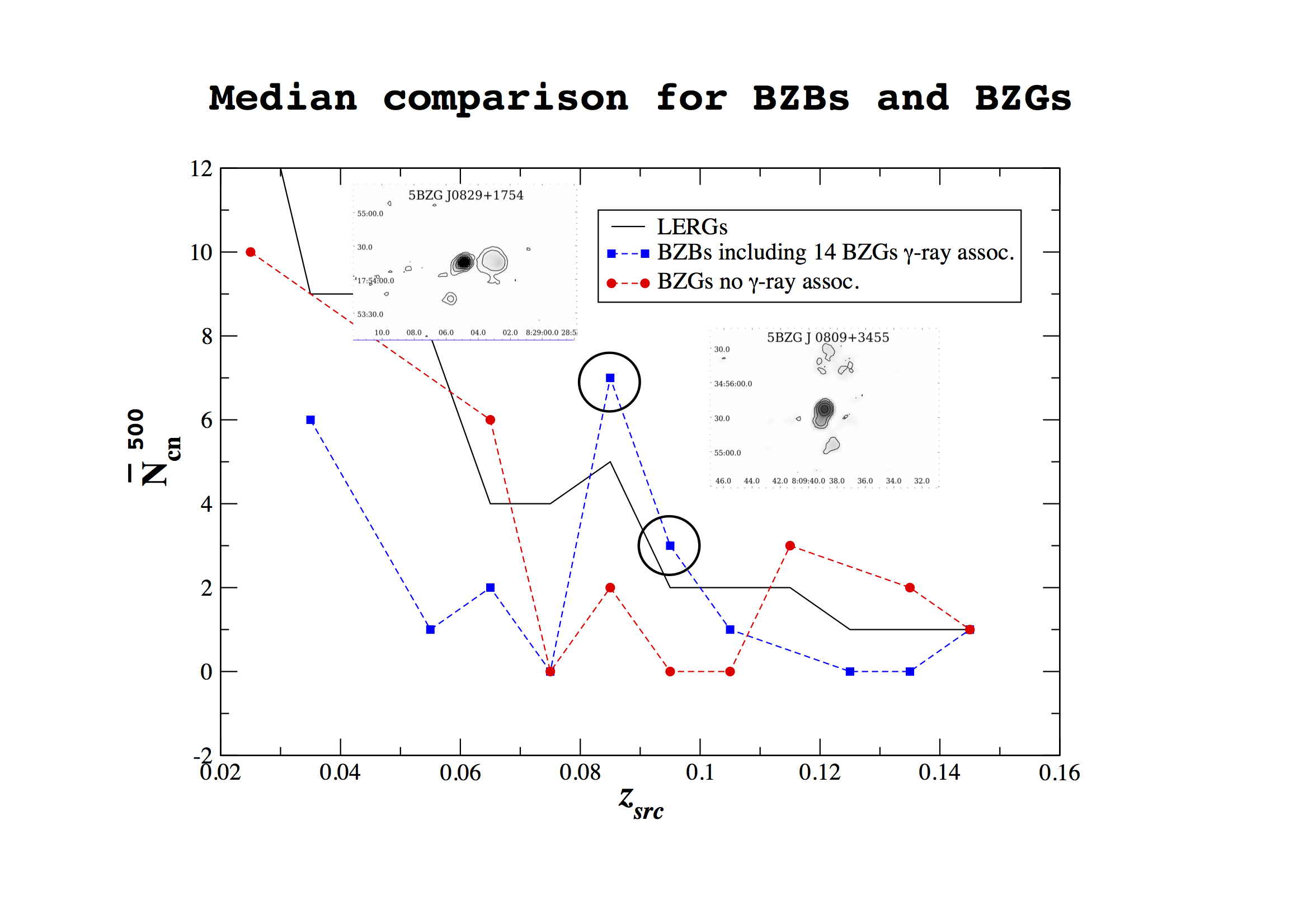}
\end{center}
\caption{Left) Same of Fig.~\ref{fig:stat} but comparing BZBs and BZGs. Here it is quite evident the agreement between BZGs and LERGs with the former class being more separated from BZBs. {\it Right)} The comparison between the BZB medians, including all BZGs associated to $\gamma$-ray sources being considered as weak BZBs, and all remaining BZGs. Two black circles highlight $z$ bins where there are mostly BZGs and that at $z=$0.075 where the two BZGs show extended radio structures highlighted with radio contour maps at 1.4 GHz 
drawn at levels of 0.0005,0.0025,0.125,0.625 Jy.}
\label{fig:bzg}
\end{figure*}
In Figure~\ref{fig:bzg} we compared BZBs and BZGs and found that values of cosmological neighbors for BZGs are consistent with those of radio galaxies. Thus mixing BZBs with BZGs, given the larger number of BZGs at $z<$0.15 (i.e., 129 in the Muriel sample) they could bias the whole statistical analysis. Moreover, Muriel (2016) that also did not consider any constraint on the ``redshift distance'' between BZBs and nearby galaxy clusters thus neglecting spurious association. 

To further explore the BZB $vs$ BZG ``dichotomy'' we assumed that all BZGs associated with a $\gamma$-ray source are ``real'' BZBs having dimmed jet emission below the host galaxy component. Then we doubled the BZB sample and we run again our comparative analysis between them and the LERGs. We found no differences with respect to the previous results with the only exception of one redshift bin between centered at 0.075 where there are only two BZGs, namely 5BZG J0809+3455 and 5BZG J0829+1754. However we inspected their FIRST radio maps and they show clear extended radio structures (i.e., lobes and plumes) beyond tens of kpc thus being very different from BZBs and being classical LERGs. 

An analysis based on similar samples used here, comparing FR\,Is and BZBs, has been recently carried out \citep[see][for details]{sandrinelli19} using the average excess of galaxy surface density $E_r$. As extensively discussed in Massaro et al (2019, 2020) this method has several statistical and cosmological uncertainties. An analysis performed without removing these biases will show that the higher redshift population tend to inhabit less galaxy-rich large-scale environments. 

\begin{figure*}[!th]
\begin{center}
\includegraphics[height=6.9cm,width=8.4cm,angle=0]{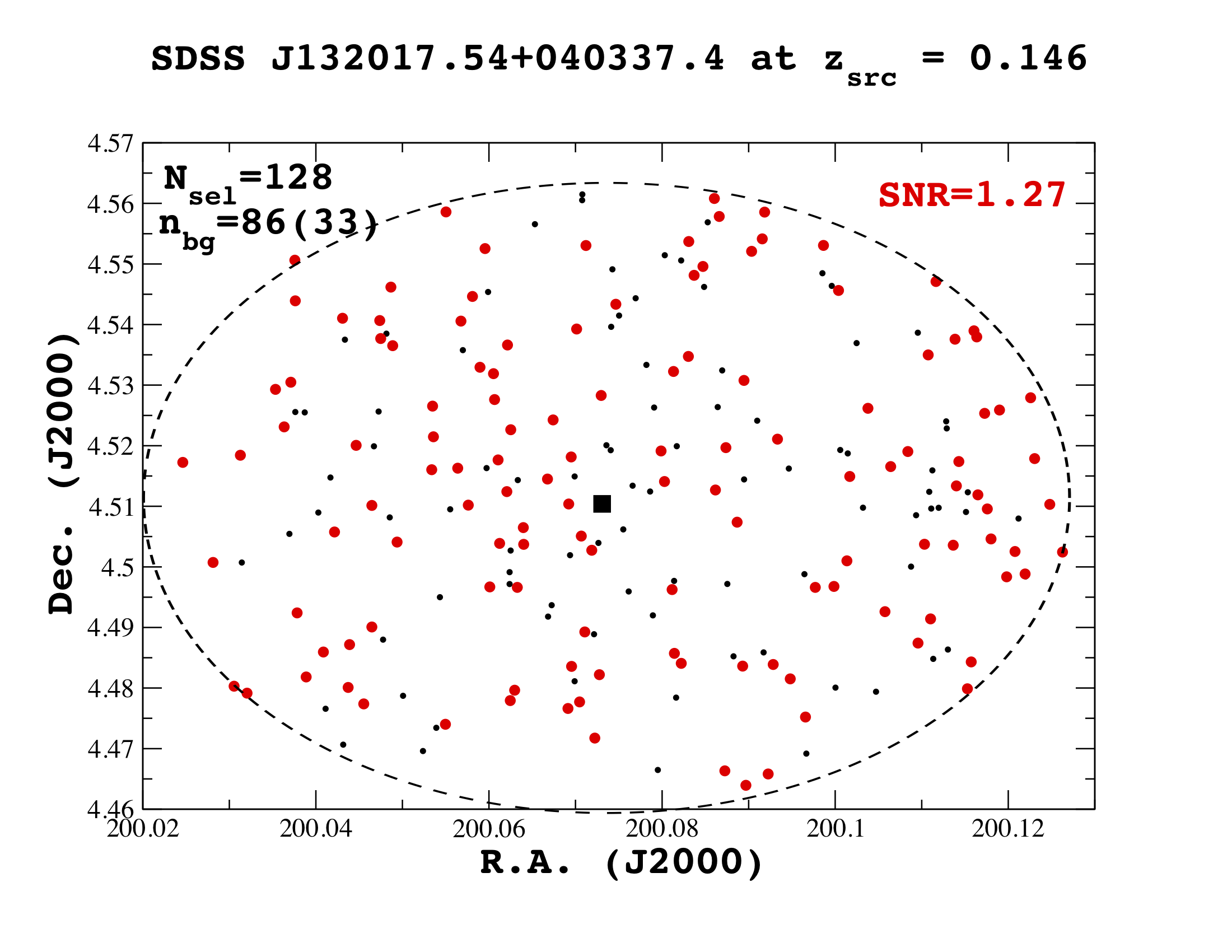}
\includegraphics[height=6.9cm,width=8.4cm,angle=0]{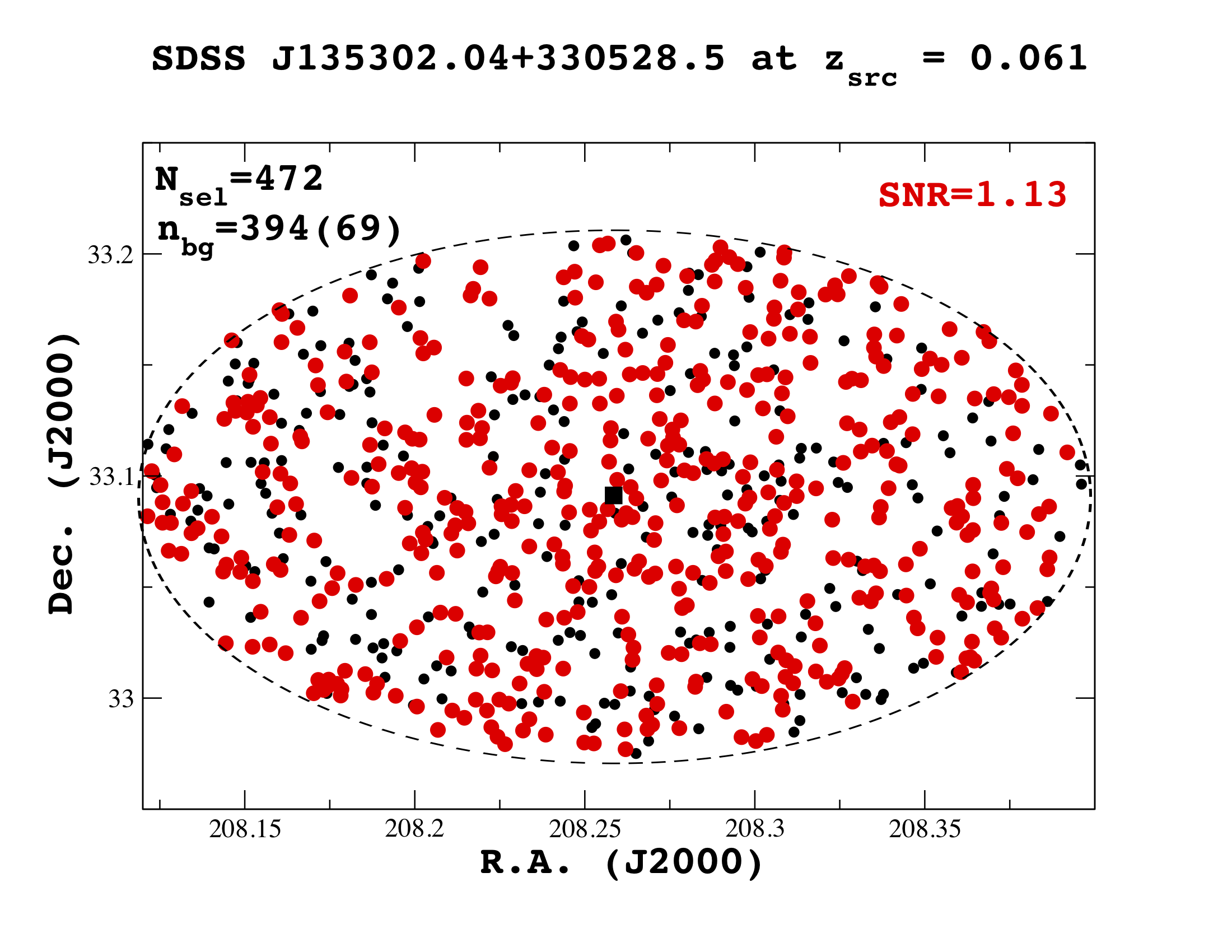}
\end{center}
\caption{All SDSS sources lying within 500\,kpc (black ellipse), computed at the redshift of two FR\,Is, lying in the center of both images: SDSS J132017.54+043037.4 (left) and SDSS J135302.04+330528.5 (right). Black circles mark sources listed as galaxies according to SDSS flags (see \S~\ref{sec:compare} for details) while those brighter than absolute magnitude in the $i$ band equal to -21, computed at the redshift of the central radio galaxy are red and their number is $N_{sel}$. The number of background galaxies $n_{bg}$ is also reported, with its standard deviation in parenthesis in both images together with the signal to noise ratio (SNR), computed averaging the galaxy count, adopting the same $i$ band magnitude selection, over 20 regions of the same area centered at angular separation grater than 4\,Mpc, from the central sources, and is reported  in Fig.~\ref{fig:snr}. Both measurements clearly show as excess of the galaxy number density $E_r=N_{sel}-n_{bg}$ is consistent with a background fluctuations at less than 2$\sigma$ (i.e., as an upper limit).}
\label{fig:snr}
\end{figure*}

Moreover this method improperly average measurements of $E_r$ with different signal-to-noise ratios (SNRs) and compare sources in different redshift bins thus including cosmological artifacts. To illustrate this SNR effect in Fig.~\ref{fig:snr} we show SDSS sources around two FR\,Is, namely SDSS J132017.54+043037.4 at $z_\mathrm{src}=$0.146 and SDSS J135302.04+330528.5 at $z_\mathrm{src}=$0.061 together with all background and foreground galaxies within a circular region of 500 kpc. Surrounding galaxies $N_{sel}$ are selected to have SDSS flags {\it q\_mode}=1, {\it Q}$>$2), {\it cl}=3 and {\it ic}=3 and an absolute magnitude in the $i$ band, computed at the same distance of the central source, greater than -21. The number of background galaxies $n_{bg}$, reported with its standard deviation in parenthesis, was estimated adopting the same criteria previously mentioned and averaging on 20 regions of the same area centered, at angular separation grater than 4\,Mpc from the central source. Both measurements clearly show as excess of galaxies is marginally significant ($\sim$1$\sigma$), thus being consistent with a background fluctuations and averaging over these measurements does not appear statistically correct. This makes challenging any comparison with our results.

\section{Summary and Conclusions}
\label{sec:summary} 
In the present analysis we focused on the comparison between large-scale environments of BZBs and radio galaxies at similar redshifts. This is the key to obtain robust results that guarantee to avoid statistical biases and cosmological artifacts affecting most of previous analyses (see \S~\ref{sec:compare} for comparison with the literature). Our analysis is carried out counting the number of cosmological neighbors, i.e., optical galaxies with a firm spectroscopic redshift and with velocities within the maximum velocity dispersion of sources belonging to galaxy groups and clusters. 

Results achieved can be summarized as follows.
\begin{enumerate}
\item In the local Universe the large-scale environment of BZBs is systematically different from that of both FR\,Is and LERGs suggesting that the unification scenario of radio-loud AGNs must be revised. 
\item However, a direct comparison between environmental properties of BZBs and the low power ``compact'' radio galaxies, known as FR\,0s, reveals that their large-scale environments are indistinguishable. This suggests that FR\,0s could be the parent population of BZBs.
\item Comparing BZBs and BZGs we also found that the latter class appears to have large-scale environments more similar to LERGs thus unlikely to be all ``weak'' BZBs. 
\item Investigations of large scale environments based on the average excess of galaxy surface density does not appear statistically robust being affect by cosmological artifacts and uncertainties due to the SNR.  
\end{enumerate}

We conclude that BZBs are mainly aligned counterparts of compact FR\,0s and only in extreme cases LERGs. This correspondence between BZBs and FR\,0s, both hosted in massive elliptical galaxies, points towards the ubiquitous presence of relativistic jets as natural outcome of gas accretion into the deep gravitational potential well produced by supermassive black holes.
\begin{table*} 
\tiny
\caption{Environmental parameters for all sample analyzed (first 10 lines).}
\label{tab:main}
\begin{center}
\begin{tabular}{lllllrrrrr}
\hline
Sample & Name  & R.A. (J2000) & Dec. (J2000) & $z_{src}$ & $\Delta\,z$ & $d_{proj}$ & $N^{500}_{cn}$ & $N^{1000}_{cn}$ & $N^{2000}_{cn}$ \\
 & & hh:mm:ss.ss & hh:mm:ss.ss & & & (kpc) & & & \\ 
\hline 
\noalign{\smallskip}
  FR0 & SDSSJ010852.48-003919.4 & 01:08:52.48 & -00:39:19.40 & 0.045 & 0.0012 & 303.05 & 2.0 & 6.0 & 7.0\\
  FR0 & SDSSJ011204.61-001442.4 & 01:12:04.61 & -00:14:42.40 & 0.044 & 4.0E-4 & 556.58 & 0.0 & 1.0 & 1.0\\
  FR0 & SDSSJ011515.78+001248.4 & 01:15:15.78 & +00:12:48.40 & 0.045 & 2.0E-4 & 173.27 & 27.0 & 39.0 & 39.0\\
  FR0 & SDSSJ015127.10-083019.3 & 01:51:27.10 & -08:30:19.30 & 0.018 & 2.0E-4 & 30.75 & 12.0 & 13.0 & 13.0\\
  FR0 & SDSSJ020835.81-083754.8 & 02:08:35.81 & -08:37:54.80 & 0.034 & 3.0E-4 & 413.79 & 1.0 & 2.0 & 2.0\\
  LERG & SDSSJ073014.37+393200.4 & 07:30:14.37 & +39:32:00.40 & 0.142 & 0.0011 & 207.56 & 1.0 & 4.0 & 6.0\\
  LERG & SDSSJ073505.25+415827.5 & 07:35:05.25 & +41:58:27.50 & 0.087 & 6.0E-4 & 758.0 & 3.0 & 4.0 & 10.0\\
  LERG & SDSSJ073719.18+292932.0 & 07:37:19.18 & +29:29:32.00 & 0.111 & 0.0034 & 826.91 & 1.0 & 1.0 & 5.0\\
  LERG & SDSSJ074125.85+480914.3 & 07:41:25.85 & +48:09:14.30 & 0.12 & 0.005 & 1890.54 & 0.0 & 0.0 & 1.0\\
  LERG & SDSSJ074351.25+282128.0 & 07:43:51.25 & +28:21:28.00 & 0.106 & 3.0E-4 & 258.8 & 4.0 & 4.0 & 9.0\\
\noalign{\smallskip}
\hline
\end{tabular}\\
\end{center}
Col. (1): Sample. \\
Col. (2): Source name. \\
Col. (3): Right Ascension.\\
Col. (4): Declination. \\
Col. (5): redshift.\\
Col. (6): Difference between the average redshift of cosmological neighbors in 2\,Mpc and that of the central source.\\
Col. (7): Physical distance between the central RG and the average position of cosmological neighbors within 2\,Mpc,  computed at $z_{src}$.\\
Col. (8,9,10): Number of cosmological neighbors within 500, 1000 ad 2000 kpc, respectively.\\
\end{table*}

\acknowledgments
F. M. wishes to thank Dr. C. C. Cheung for their valuable discussions on this project planned during the organization of the IAU 313 Symposium. 
This work is supported by the ``Departments of Excellence 2018 - 2022'' Grant awarded by the Italian Ministry of Education, University and Research (MIUR) (L. 232/2016). This research made use of resources provided by the Ministry of Education, Universities and Research for the grant MASF\_FFABR\_17\_01. This investigation is supported by the National Aeronautics and Space Administration (NASA) grants GO9-20083X.
TOPCAT and STILTS astronomical software \citep{taylor05} were used for the preparation and manipulation of the tabular data and the images.

~

\end{document}